\documentclass[aps,preprint,superscriptaddress,12pt]{revtex4}%
\usepackage{amssymb}
\usepackage{amsfonts}
\usepackage{amsmath}
\usepackage{graphicx}
\usepackage[colorlinks,hyperindex]{hyperref}%
\providecommand{\U}[1]{\protect\rule{.1in}{.1in}}
\hypersetup{
colorlinks,
citecolor=blue,
linkcolor=black,
urlcolor=black,
}

\begin{document}
\title{On the Hamilton-Jacobi method in classical and quantum nonconservative systems}
\author{A. de Souza Dutra}
\email{dutra@feg.unesp.br}
\affiliation{UNESP, Universidade Estadual Paulista, Campus de Guaratinguet\'{a}, 12516-410,
SP, Brazil}
\author{R. A. C. Correa}
\email{rafael.couceiro@ufabc.edu.br}
\affiliation{\ CCNH, Universidade Federal do ABC, 09210-580, Santo Andr\'{e}, SP, Brazil}
\author{P. H. R. S. Moraes}
\email{moraes.phrs@gmail.com}
\affiliation{ITA, Instituto Tecnol\'{o}gico de Aeronautica, 12228-900, S\~{a}o Jos\'{e} dos
Campos, SP, Brazil}

\begin{abstract}
In this work we show how to complete some Hamilton-Jacobi solutions of linear,
nonconservative classical oscillatory systems which appeared in the literature
and we extend these complete solutions to the quantum mechanical case. In
addition, we get the solution of the quantum Hamilton-Jacobi equation for an
electric charge in an oscillating pulsing magnetic field. We also argue that
for the case where a charged particle is under the action of an oscillating
magnetic field, one can apply nuclear magnetic resonance techniques in order
to find experimental results regarding this problem. We obtain all results
analytically, showing that the quantum Hamilton-Jacobi formalism is a powerful
tool to describe quantum mechanics.

\end{abstract}
\maketitle


\section{Introduction}

In 1924 the physicist Max Born put forward for the first time the name
\textquotedblleft quantum mechanics\textquotedblright\ in the literature
\cite{born}. In that work, quantum mechanics denoted a theoretical framework
of atomic and electronic motion, which was understood in the same level of
generality and consistency of the classical mechanics laws. Approximately one
year after that work, in 1925, the historical paper presented by Heisenberg
and entitled \textquotedblleft Quantum-theoretical reinterpretation of
kinematic and mechanical relations\textquotedblright\cite{hein} has shown a
new quantum-theoretical quantity which contains information about the
measurable line spectrum of an atom. Motivated by Heisenberg's work, Born,
Jordan and Heisenberg published the articles \textquotedblleft On quantum
mechanics\textquotedblright\ \cite{jordan} and \textquotedblleft On quantum
mechanics II\textquotedblright\ \cite{jordan1}, which were the first
comprehensive explanations of quantum mechanics. It is worth mentioning that
those works have been performed using the matrix framework.

On the other hand, Dirac formulated independently a consistent algebraic
framework of quantum mechanics \cite{dirac}, where the equations were obtained
with no use of matrix theory.

However, it was only in 1926 that the Schr\"{o}dinger formalism (SF) appeared
in the literature. Since then, day after day, several problems linked to
quantum mechanics have been analyzed rigorously in the literature \cite{d1,
d2, d3, d4, d5}. Formal developments have been arisen, in particular to deepen
the comprehension regarding quantum fields. Quantum canonical transformations
have attracted interest since the incipient development of the theory about
one century ago.

Although the SF is a prevailing framework, alternative formalisms emerged. For
instance, the path integral formulation plays a prominent role in quantum
field theory \cite{b1}.

The basic postulates of a third version for the study of quantum mechanics
have also been proposed, namely a quantum version of the Hamilton-Jacobi
formalism \cite{b7}, where a better understanding of the quantum
HamiIton-Jacobi theory and its consequences was presented. Moreover, in that
work the authors have shown applications of the quantum Hamilton-Jacobi
formalism (QHJF) for the calculation of the propagators of the harmonic
oscillator potential and of the same potential with time-dependent parameters.
Here, it is important to highlight that Leacock and Padgett (LP) \cite{page}
and independently Gozzi \cite{gozzi} are a few names who have worked this
formalism out. For instance, LP developed the QHJF for the case of
conservative systems, where the main feature of their theory is the definition
of the quantum action variable which permits the determination of the
bound-state energy levels without solving the dynamic equation \cite{page}. On
the other hand, Castro and Dutra (CD) have obtained the QHJF through basic
postulates similar to the case of the Heisenberg picture \cite{b7}. An
important feature in CD's work is the straightforward equivalence of the QHJF
with both the Feynman and Schr\"{o}dinger formalisms.

Currently we can find in several areas of physics a considerable amount of
works dedicated to the studies of the QHJF. Among the different research
areas, we can find an interesting connection of quantum Hamilton--Jacobi
theory with supersymmetric quantum mechanics (SUSYQM) \cite{susy,rasina}. In
this case, the quantum momenta of supersymmetric partner potentials are
connected via linear fractional transformations. Moreover, in the SUSYQM
context, it has been shown by Dauod and Kibler a connection between fractional
and ordinary SUSYQM \cite{dauod}. Another line of investigation comes from
one-dimensional scattering problems in the framework of the QHJF \cite{wyatt}.
In addition, Roncadelli and Schulman solved the quantum Hamilton-Jacobi
equation, by a prescription based upon the propagator of the Schr\"{o}dinger
equation \cite{ronca}. It provided the use of quantum Hamilton-Jacobi theory,
developing an unexpected relation between operator ordering and the density of
paths around a semiclassical trajectory. Related to it, black hole tunnelling
procedures have been placed as prominent methods to calculate the temperature
of black holes using the Hamilton Jacobi technique in the Wentzel, Kramers,
and Brillouin (WKB) approximation \cite{1,2,Vanzo:2011wq}. Various types of
black holes have been studied in the context of tunnelling of fermions and
bosons as well \cite{1,2,Vanzo:2011wq,meuepl}. Tunnelling procedures are quite
well used to investigate black holes radiation, by taking into account
classically forbidden paths that particles go through, from the inside to the
outside of black holes. Moreover, quantum WKB approaches were employed to
calculate corrections to the Bekenstein-Hawking entropy for the Schwarzschild
black hole \cite{d6}.

As we can see in \cite{yang}, the problem of the electron quantum dynamics in
hydrogen atom has been modeled exactly by QHJF, where the quantization of
energy, angular momentum, and the action variable are originated from the
electron complex motion. In addition, the shell structure observed in hydrogen
atom arises from the structure of the complex quantum potential, from which
the quantum forces acting upon the electron can be uniquely determined.

Moreover, much has been learned regarding the QHJF in the last years, when
several developments have been accomplished in the literature. These include
the definability of time parameterization of trajectories \cite{adc1},
corrections for any soliton equation for which action-angle variables are
known \cite{adc2}, lattice theories \cite{adc3}, gauge invariance in loop
quantum cosmology \cite{adc4}, treatment of the relativistic double
ring-shaped Kratzer potential \cite{adc5}, shape invariant potentials in
higher dimensions \cite{adc6}, application to the photodissociation dynamics
of NOCl \cite{adc7}, and Dirac-Klein-Gordon systems \cite{adc8}.

Furthermore, Vujanovic and Strauss \cite{b8} developed a series of
calculations using the classical Hamilton-Jacobi method to study linear
nonconservative systems. In order to obtain solutions for the cases studied,
the authors used an expression for the classical action that contains only the
quadratic term, which reads:
\begin{equation}
S_{VS}(x,t)=\frac{\alpha(t)}{2}x^{2}\,. \label{1}%
\end{equation}
Despite this term does not alter the classical solution, here we shall show
that it does not hold for the quantum mechanical case. In fact, when quantum
systems are approached, we shall study the Hamilton's principal function $S$
given by a polynomial of $x$, which is written in the form
\begin{equation}
S(x,t)=\frac{\alpha(t)x^{2}}{2}+\xi(t)x+\zeta(t)\,. \label{2}%
\end{equation}
In fact, the linear term is necessary for the development of the quantum
propagator. Hence, this term can not be neglected when quantum solutions are
regarded. In addition, in order to deal with a more interesting application
from the point of view of QHJF, we will study the problem of an electric
charge in an oscillating pulsed magnetic field \cite{B14, B13}.

This paper is organized as follows. In the next section, we present a complete
review about the QHJF and its basic postulates. In Sec. \textcolor{red}{III},
we show an illustration of the QHJF to the standard case of the harmonic
oscillator. In Sec. \textcolor{red}{IV}, we apply the ideas to analyze the
driven oscillator case. Section \textcolor{red}{V} is devoted to the resonance
example. In Sec. \textcolor{red}{VI}, we show an application of the
Hamilton-Jacobi formalism to the problem related with the quantum dynamics of
an electric charge in an oscillating pulsing magnetic field. We end up with
some general remarks and conclusions in Sec. \textcolor{red}{VII}.

\section{A brief review on Hamilton-Jacobi formalism}

In this section we will present a review about the QHJF and its basic
postulates. We present a prescription for obtaining the QHJE from the
classical one. At this point, it is important to remark that this approach is
analogous to the Heisenberg prescription, which makes a link between the
Poisson brackets and quantum commutation relations. Here, we follow the work
presented by CD \cite{b7}, and revisit the QHJF as well.

Let us start by remembering that the Hamilton principal function, or action,
$S_{cl}$, is a generating function of the canonical transformation $(\vec
{r},\vec{p})\mapsto(\vec{r}\prime,\vec{p}\prime)$, which generates new
time-dependent variables $\vec{r}\prime$ and $\vec{p}\prime$ with null
Hamiltonian. In this case, the classical Hamilton-Jacobi equation reads
\begin{equation}
H(\vec{r},\vec{\nabla}S_{cl},t)+\frac{\partial S_{cl}}{\partial t}=0,
\label{d1}%
\end{equation}
where $\vec{\nabla}S_{cl}=\vec{p}$. It is worthwhile to point out that the
above classical Hamilton-Jacobi equation provides a successful form for
establishing the equations of motion of a mechanical system.

Following the approach given in \cite{b7}, where the authors used classical
mechanics as a short wavelength limit of wave mechanics, and by taking into
consideration the similarity with the electromagnetic quantities and their
limits to geometrical optics, it was postulated that the quantum wave
amplitude has the form
\begin{equation}
\Psi=2^{-1/2}\exp\left(  iS/\hslash\right)  , \label{d2}%
\end{equation}

\noindent where $S$ is the quantum Hamilton's principal function, or complex
action, $\hslash$ represents the Planck constant and $2^{-1/2}$ is a factor
introduced for convenience. Therefore, the action $S$ can be realized as a
phase of the wave motion process. In order to accomplish the transition from
the classical Hamilton-Jacobi equation to the quantum case, one defines the
momentum in the operatorial form, given by%
\begin{equation}
\vec{p}_{op}=\vec{\nabla}S-{i\hslash}\vec{\nabla}. \label{d3}%
\end{equation}
Hence the classical momentum is obtained in the limit $\hslash\rightarrow0$,
where the commutation relations are established. Thus, when the Hamiltonian
has the standard form%
\begin{equation}
H=\frac{\vec{p}^{2}}{2m}+V, \label{d4}%
\end{equation}
\noindent one can find, using (\ref{d1}) and (\ref{d3}), the following quantum
Hamilton-Jacobi equation (QHJE):
\begin{equation}
\frac{1}{2m}(\vec{\nabla}S)^{2}+\frac{\partial S_{cl}}{\partial t}%
+V=\frac{i\hslash}{2m}\nabla^{2}S. \label{d5}%
\end{equation}

In the next sections we will show how linear, strictly nonconservative,
oscillatory systems with one degree of freedom may be analyzed within the
quantum Hamilton-Jacobi framework. The motivation for this study is that
linear dissipative systems, possessing even one degree of freedom, have not
been analyzed in the context of the quantum Hamilton-Jacobi method, despite of
its practical, theoretical, and pedagogical interests.

\section{\smallskip Harmonic oscillator}

A particular important physical system is the harmonic oscillator. There
exists a large number of important physical applications for it, such as the
vibrations of the atoms of a molecule about their equilibrium position or even
an electromagnetic field, for instance. In fact, whenever the behavior of a
physical system in the neighborhood of a stable equilibrium position is
studied, one obtains equations which, in the limit of small oscillations, are
those of a harmonic oscillator.

Let us start our study with a straightforward example of the harmonic
oscillator. The associated quantum Hamilton-Jacobi equation is provided by
\cite{b7}
\begin{equation}
\frac{\partial S}{\partial t}+\frac{1}{2}\left(  \frac{\partial S}{\partial
x}\right)  ^{2}+\frac{\omega^{2}x^{2}}{2}=\frac{i\hbar}{2}\frac{\partial^{2}%
S}{\partial x^{2}}. \label{3}%
\end{equation}
\smallskip

The substitution of \smallskip(\ref{2}) \noindent into the QHJE (\ref{3})
generates a polynomial equation leading to a system of first-order coupled
differential equations for the arbitrary coefficients introduced in (\ref{2}).
The polynomial equations can be split into the following set of first-order
non-linear differential equations:
\begin{align}
\dot{\alpha}(t)+\alpha^{2}(t)+\omega^{2} &  =0,\nonumber\\
\dot{\xi}(t)+\alpha(t)\xi(t) &  =0,\nonumber\\
\dot{\zeta}(t)+\frac{\xi^{2}(t)}{2}-\frac{i\hbar}{2}\alpha(t) &  =0,\label{4}%
\end{align}
\smallskip\noindent\noindent\noindent yielding the general solutions
\begin{align}
\alpha(t) &  =-\omega\tan(\omega t+c_{1}),\nonumber\\
\xi(t) &  =c_{2}\sec(\omega t+c_{1}),\nonumber\\
\zeta(t) &  =-\frac{c_{2}^{2}}{2\omega}\tan(\omega t+c_{1})+\frac{i\hbar}%
{2}\ln[\cos(\omega t+c_{1})]+c_{3},\label{9}%
\end{align}
where $c_{1}$, $c_{2\text{ }}$ and $c_{3}$ are arbitrary integration
constants. Hence a complete solution of (\ref{1}) is given by
\begin{equation}
\!\!\!\!\!\left.  S(x,t)=-\frac{\omega}{2}\tan(\omega t+c_{1})x^{2}+c_{2}%
\sec(\omega t+c_{1})x-\frac{c_{2}^{2}}{2\omega}\tan(\omega t+c_{1}%
)+\frac{i\hbar}{2}\ln[\cos(\omega t+c_{1})]+c_{3}.\right.  \label{cm1}%
\end{equation}
\smallskip It is worth to emphasize that in the limit $\hbar\rightarrow0$, the
classical Hamilton's principal function is reobtained. The general solution
for the classical case of the Hamilton-Jacobi equation can be obtained from
the constraint $\frac{\partial S}{\partial c_{1}}=B$, where $B$ is a constant.
Furthermore, it is straightforward to verify that the classical solution is
given by
\begin{equation}
x_{\pm}(t)=\frac{c_{2}\sin(\omega_{0}t+c_{1})}{\omega_{0}}\pm\left[
-\frac{c_{2}^{2}}{\omega_{0}^{2}}-\frac{2B}{\omega_{0}}\right]  ^{1/2}%
\!\!\!\!\!\!\cos(\omega_{0}t+c_{1}).\label{11}%
\end{equation}
\noindent By analyzing the classical case for Eq.(\ref{cm1}), the solution can
also be immediately determined by $\frac{\partial S}{\partial c_{2}}=B$. Thus,
in this case the classical solution contains two integration constants, as it
should be expected, since the equation of motion is a second-order one.

Moreover, by using Eq.(\ref{cm1}), the solution for the problem consists in
obtaining the quantum propagator, by imposing the following boundary condition
\cite{b7}
\begin{equation}
S(x,0)=\hbar kx. \label{13}%
\end{equation}
\smallskip\noindent Therefore $c_{1}=0$, $c_{2}=\hbar k$ and $c_{3}=0$.

The concept of propagators is of great importance in quantum physics and in
the Feynman's formulation, particularly. All the time evolution of a given
system may be obtained through the propagators \cite{b7}. They are used mostly
to calculate the probability amplitude for particle interactions using Feynman diagrams.

The propagator can be obtained by considering a physical wave packet
\begin{equation}
\Psi(x,t)=\frac{1}{\sqrt{2\pi}}\int dk\Phi(k)\exp\left[  \frac{i}{\hbar}%
S_{k}(x,t)\right]  , \label{50.2}%
\end{equation}
\noindent where $S_{k}(x,t)$ denotes the quantum Hamilton's principal function
if the boundary condition $S_{k}(x,0)=\hbar kx$ is taken into account.
Inserting the Fourier transform $\Phi(k)=\frac{1}{\sqrt{2\pi}}\int
dx\Psi(x,0)\exp(-ikx)$ in Eq.(\ref{50.2}) yields
\begin{equation}
\Psi(x,t)=\int dkK(x,t;\tilde{x},0)\Psi(\tilde{x},0)\,,
\end{equation}
\noindent where the propagator reads
\begin{equation}
K(x,t,\tilde{x},0)=\frac{1}{2\pi}\int dk\exp\left\{  \frac{i}{\hbar}\left[
S(x,t)-\hbar k\tilde{x}\right]  \right\}  . \label{14}%
\end{equation}
\smallskip We observe that the constant $c_{2\text{ }}$ is related to the term
that generates the quantum propagator. It is important to remark that this
constant appears in the linear term of Eq.(\ref{2}). Hence we conclude that
the linear term must also compose the principal Hamilton function, in order to
construct the quantum propagator.

By substituting the solution and the initial conditions imposed to the
expression of the propagator and integrating in $k$, one gets
\begin{equation}
K(x,t,\tilde{x},0)=\left(  \frac{\omega}{2\pi i\hbar\sin(\omega t)}\right)
^{1/2}\exp\left\{  \frac{i\omega}{2\hbar\sin(\omega t)}\left[  (x^{2}%
+\tilde{x}^{2})\cos(\omega t)-2x\tilde{x}\right]  \right\}  . \label{15}%
\end{equation}

The quantum propagator can be alternatively constructed \cite{b7}, by imposing
that
\begin{equation}
K(x,t;\tilde{x},0)=\exp\left[  \frac{i}{\hbar}S(x,t;\tilde{x},0)\right]
,\label{16}%
\end{equation}
\smallskip\noindent where $S$ represents the quantum solution of the equation
of Hamilton-Jacobi.

The propagator must satisfy the condition
\begin{equation}
\lim_{t\rightarrow0^{+}}K(x,t;\tilde{x},0)=\delta(x-\tilde{x}), \label{17}%
\end{equation}
\smallskip\noindent where $\delta(x-\tilde{x})$ represents the Dirac delta
function. For our purposes it is useful to employ the following
representation:
\begin{equation}
\delta(x-\tilde{x})=\lim_{t\rightarrow0^{+}}(\pi\lambda t)^{-\frac{1}{2}}%
\exp\left[  -\frac{(x-\tilde{x})^{2}}{\lambda t}\right]  . \label{18}%
\end{equation}
\smallskip By using Eqs.(\ref{16} - \ref{18}), we determine
\begin{equation}
c_{1}=\frac{\pi}{2},\qquad c_{2}=\omega\tilde{x},\qquad c_{3}=-i\frac{\hbar
}{2}\ln\left(  \frac{i\omega}{2\pi\hbar}\right)  . \label{19}%
\end{equation}
\smallskip By substituting Eq.(\ref{19}) in Eq.(\ref{16}), the propagator is
reduced to the form presented in (\ref{15}).

We emphasize that the linear term in $S$ is quite necessary. Hereon we are
going to implement this approach in similar cases which, up to our knowledge,
have not been taken into account in the literature, at least from the point of
view of the quantum Hamilton-Jacobi formalism.

\section{Driven oscillator}

Driven harmonic oscillators are damped oscillators further affected by an externally applied force. The potential of a driven harmonic oscillator can describe many phenomena in physics, such as superconducting quantum-interference devices \cite{rose-innes/1978} and magnetohydrodynamics \cite{king/1983}.

Its classical equation of motion reads
\begin{equation}
\ddot{x}+\omega x^{2}=h\cos(\Omega t), \label{22}%
\end{equation}
\noindent and the corresponding Lagrangian can be written as
\begin{equation}
L=\frac{1}{2}\left[  \left(  \dot{x}-\dot{f}{(}t)\right)  ^{2}-\omega
^{2}\left(  x-f(t)\right)  ^{2}\right]  , \label{23}%
\end{equation}
\noindent where $f(t)=\left(  \frac{h\cos(\Omega t)}{\omega^{2}-\Omega^{2}%
}\right)  .$ The following Hamiltonian is then derived:
\begin{equation}
H=\frac{p^{2}}{2}+\dot{f}{(}t)p+\frac{\omega^{2}}{2}[x-f(t)]^{2}. \label{24}%
\end{equation}
Hence, the Hamilton-Jacobi equation assumes the form
\begin{equation}
\frac{\partial S}{\partial t}+\frac{1}{2}\left(  \frac{\partial S}{\partial
x}\right)  ^{2}-\left[  \frac{h\Omega\sin(\Omega t)}{\omega^{2}-\Omega^{2}%
}\right]  \frac{\partial S}{\partial x}+\frac{1}{2}\omega^{2}\left[  x-\left(
\frac{h\cos(\Omega t)}{\omega^{2}-\Omega^{2}}\right)  \right]  ^{2}%
=\frac{i\hbar}{2}\frac{\partial^{2}S}{\partial x^{2}}, \label{25}%
\end{equation}
\smallskip\noindent and the principal Hamilton function is represented by
\begin{equation}
S(x,t)=\frac{1}{2}\alpha(t)\left[  x-f(t)\right]  ^{2}+\xi(t)\left[
x-f(t)\right]  +\zeta(t). \label{26}%
\end{equation}
\smallskip\noindent By substituting Eq.(\ref{26}) in Eq.(\ref{25}), the
quantum Hamilton's principal function reads
\begin{align}
S(x,t)  &  =-\frac{1}{2}\omega\tan(\omega t+c_{1})\left[  x-\left(
\frac{h\cos(\Omega t)}{\omega^{2}-\Omega^{2}}\right)  \right]  ^{2}+c_{2}%
\sec(\omega t+c_{1})\left[  x-\left(  \frac{h\cos(\Omega t)}{\omega^{2}%
-\Omega^{2}}\right)  \right] \nonumber\\
&  -\frac{c_{2}^{2}}{2\omega}\tan(\omega t+c_{1})+\frac{i\hbar}{2}\ln\left[
\cos(\omega t+c_{1})\right]  +c_{3}. \label{30}%
\end{align}
\smallskip The limit $\hbar\rightarrow0$ leads to the classical case, and the
solution is obtained by imposing that $\frac{\partial S}{\partial c_{1}}=B$,
implying that
\begin{equation}
x_{\pm}(t)=\frac{c_{2}\sin(\omega t+c_{1})}{\omega}\pm\left[  -\frac{c_{2}%
^{2}}{\omega^{2}}-\frac{2B}{\omega}\right]  ^{1/2}\cos(\omega t+c_{1})+\left(
\frac{h\cos(\Omega t)}{\omega^{2}-\Omega^{2}}\right)  . \label{31}%
\end{equation}
\smallskip\noindent Our result can be led to the one in \cite{b8}, with some
mathematical manipulations. The above solution can also be obtained by
imposing \smallskip$\frac{\partial S}{\partial c_{2}}=B$.

For the quantum case, once again the condition
\begin{equation}
S(x,0)=\hbar kx \label{33}%
\end{equation}
\smallskip shall be imposed, what implies that
\begin{equation}
c_{1}=0,\qquad c_{2}=\hbar k,\qquad c_{3}=\frac{\hbar kh}{\omega^{2}%
-\Omega^{2}}. \label{34}%
\end{equation}
\noindent Remembering that $f(t)=\left(  \frac{h\cos(\Omega t)}{\omega
^{2}-\Omega^{2}}\right)  $, and imposing the described conditions in
(\ref{33}) and (\ref{34}), the propagator reads
\begin{align}
&  \left.  K(x,t,\tilde{x},0)=\left(  \frac{\omega}{2\pi i\hbar\sin(\omega
t)}\right)  ^{1/2}\exp\left\{  \frac{i\omega}{2\hbar\sin(\omega t)}\left[
\left(  \left(  x-\frac{h\cos(\Omega t)}{\omega^{2}-\Omega^{2}}\right)
^{2}\right.  \right.  \right.  \right. \nonumber\\
&  \left.  \left.  \left.  \left.  +\left(  \tilde{x}-\frac{h\cos(\Omega
t)}{\omega^{2}-\Omega^{2}}\right)  ^{2}\right)  \cos(\omega t)-2\left(
x-\frac{h\cos(\Omega t)}{\omega^{2}-\Omega^{2}}\right)  \left(  \tilde
{x}-\frac{h\cos(\Omega t)}{\omega^{2}-\Omega^{2}}\right)  \right]  \right\}
.\right.  \label{37}%
\end{align}

\smallskip\smallskip From the initial condition of the second method, the
propagator can be obtained if we choose
\begin{equation}
c_{1}=\frac{\pi}{2},\qquad c_{2}=\omega\left(  \tilde{x}-f(t)\right)  ,\qquad
c_{3}=-i\frac{\hbar}{2}\ln\left(  \frac{i\omega}{2\pi\hbar}\right) ,
\label{38}%
\end{equation}
\noindent which lead to the result in (\ref{37}).

\section{ Resonances}

Resonance occurs when a given system is driven to oscillate by another vibrating system with greater amplitude at a specific preferential frequency. They occur with all types of waves, such as mechanical, electromagnetic and quantum wave functions.

Let us consider the following equation:
\begin{equation}
\overset{}{\ddot{x}}+\omega^{2}x=h\cos(\omega t). \label{41}%
\end{equation}
\smallskip\noindent The Lagrangian reads
\begin{equation}
L=\frac{1}{2}\left[  \overset{}{\dot{x}}-\frac{1}{2}ht\cos(\omega t)-\frac
{h}{2\omega}\sin(\omega t)\right]  ^{2}-\frac{1}{2}\omega^{2}\left[
x-\frac{ht}{2\omega}\sin(\omega t)\right]  , \label{42}%
\end{equation}
\smallskip\noindent whereas the Hamiltonian is given by
\begin{equation}
H=\frac{1}{2}p^{2}+\left[  \frac{1}{2}ht\cos(\omega t)+\frac{ht}{2\omega}%
\sin(\omega t)\right]  p+\frac{1}{2}\omega^{2}\left[  x-\frac{ht}{2\omega}%
\sin(\omega t)\right]  ^{2}. \label{43}%
\end{equation}
\smallskip Hence the corresponding quantum Hamilton-Jacobi equation becomes
\begin{equation}
\frac{\partial S}{\partial t}+\frac{1}{2}\left(  \frac{\partial S}{\partial
x}\right)  ^{2}+f\dot{(}t)\left(  \frac{\partial S}{\partial x}\right)
+\frac{1}{2}\omega^{2}\left[  x-f(t)\right]  ^{2}=\frac{i\hbar}{2}\left(
\frac{\partial^{2}S}{\partial x^{2}}\right)  . \label{44}%
\end{equation}
\smallskip\noindent Considering the Eq.(\ref{26})\smallskip, \noindent where
$f(t)=\left(  \frac{ht}{2\omega}\sin(\omega t)\right)  $, and applying it in
Eq.(\ref{44}), it follows that
\begin{align}
S(x,t)  &  =-\frac{1}{2}\omega\tan(\omega t+c_{1})\left[  x-\left(  \frac
{ht}{2\omega}\sin(\omega t)\right)  \right]  ^{2}+c_{2}\sec(\omega
t+c_{1})\left[  x\right. \nonumber\\
&  \left.  -\left(  \frac{ht}{2\omega}\sin(\omega t)\right)  \right]
-\frac{c_{2}^{2}}{2\omega}\tan(\omega t+c_{1})+\frac{i\hbar}{2}\ln\left[
\cos(\omega t+c_{1})\right]  +c_{3}. \label{46}%
\end{align}
In the classical case we have the solutions
\begin{equation}
x_{\pm}(t)=\frac{c_{2}\sin(\omega t+c_{1})}{\omega}\pm\left[  -\frac{c_{2}%
^{2}}{\omega^{2}}-\frac{2B}{\omega}\right]  ^{1/2}\cos(\omega t+c_{1})+\left(
\frac{ht}{2\omega}\sin(\omega t)\right)  . \label{47}%
\end{equation}
\noindent By imposing the condition
\begin{equation}
S(x,0)=\hbar kx-\hbar kf(t), \label{49}%
\end{equation}
\smallskip\noindent the quantum propagator for the resonance reads
\begin{align}
&  \left.  K(x,t;\tilde{x},0)=\left(  \frac{\omega}{2\pi i\hbar\sin(\omega
t)}\right)  ^{1/2}\exp\left\{  \frac{i\omega}{2\hbar\sin(\omega t)}\left[
\left(  \left(  x-\frac{ht}{2\omega}\sin(\omega t)\right)  ^{2}+\right.
\right.  \right.  \right. \nonumber\\
&  \left.  \left.  \left.  \left.  +\,\left(  \tilde{x}-\frac{ht}{2\omega}%
\sin(\omega t)\right)  ^{2}\right)  \cos(\omega t)-2\left(  x-\frac
{ht}{2\omega}\sin(\omega t)\right)  \left(  \,\tilde{x}-\frac{ht}{2\omega}%
\sin(\omega t)\right)  \right]  \right\}  .\right.  \label{50.1}%
\end{align}
On the other hand, if we try to construct the propagator from the initial
conditions procedure, we find
\begin{equation}
c_{1=}\frac{\pi}{2},\quad c_{2}=\omega\left[  \tilde{x}-f(t)\right]  ,\quad
c_{3}=-i\frac{\hbar}{2}\ln\left(  \frac{i\omega}{2\pi\hbar}\right)  .
\label{51}%
\end{equation}
\smallskip With these values, the propagator (\ref{16}) is led to the form
given by Eq.(\ref{50.1}).

\section{Electric charge in an oscillating pulsed magnetic field}

In this section we show an application of the Hamilton-Jacobi formalism to the
problem related with the quantum dynamics of an electric charge in an
oscillating pulsed magnetic field \cite{B14}. It becomes important then to
analyze, through a parallel formalism, the validity of the solutions
presented, since the systems can describe experimental measurements in nuclear
magnetic resonance techniques \cite{B13}.

We consider an electric charge $e$ in an oscillating pulsed magnetic field
given by
\begin{equation}
\vec{B}(t)=B_{1}\cos(\omega t)\hat{\imath}+B_{2}\sin(\omega t)\hat{\jmath
}+B_{0}\hat{k}\,. \label{p1}%
\end{equation}
\noindent The Lagrangian for a charge in an electromagnetic field reads
\begin{equation}
L=\frac{m\vec{v}^{2}}{2}+\frac{e}{c}\vec{A}\cdot\vec{v}-e\phi(\vec{r})\,,
\label{p2}%
\end{equation}
where $\vec{A}=-\frac{1}{2}\left(  \vec{r}\times\vec{B}\right)  $ and
$\phi(\vec{r})$ denotes the scalar potential. The Hamiltonian is usually
written as
\begin{equation}
H=\frac{1}{2m}\left(  \vec{p}-\frac{e}{c}\vec{A}\right)  ^{2}+e\phi(\vec
{r})\,, \label{p4}%
\end{equation}
or explicitly, as
\begin{align}
&  \left.  H=\frac{p_{x}^{2}+p_{y}^{2}+p_{z}^{2}}{2m}+\frac{m\gamma^{2}%
B_{0}^{2}}{8c^{2}}(x^{2}+y^{2})+\frac{m\gamma^{2}B_{1}^{2}}{8c^{2}}\left[
z^{2}+(x\sin(\omega t)-y\cos(\omega t))^{2}\right]  \right. \nonumber\\
&  \left.  -\frac{m\gamma^{2}B_{0}B_{1}}{4c^{2}}z(y\sin(\omega t)-x\cos(\omega
t))+\frac{\gamma B_{1}}{2c}\cos(\omega t)L_{x}-\frac{\gamma B_{1}}{2c}%
\sin(\omega t)L_{y}+\frac{\gamma B_{0}}{2c}L_{z}+e\phi(\vec{r}),\right.
\label{p11}%
\end{align}
\noindent where $\gamma\equiv\frac{e}{m}$. By substituting $\vec{p}%
=\vec{\nabla}S+\frac{\hbar}{i}$ and $H=-\frac{\partial S}{\partial t}$,
yields
\begin{align}
&  \left.  \frac{1}{2m}\left(  \vec{\nabla}S\right)  ^{2}+\frac{m\gamma
^{2}B_{0}^{2}}{8c^{2}}(x^{2}+y^{2})+\frac{m\gamma^{2}B_{1}^{2}}{8c^{2}}\left[
z^{2}+(x\sin(\omega t)-y\cos(\omega t))^{2}\right]  \right. \nonumber\\
&  \left.  -\frac{m\gamma^{2}B_{0}B_{1}}{4c^{2}}z(y\sin(\omega t)-x\cos(\omega
t))+\frac{\gamma B_{1}}{2c}\cos(\omega t)L_{x}-\frac{\gamma B_{1}}{2c}%
\sin(\omega t)L_{y}\right. \nonumber\\
&  \left.  +\frac{\gamma B_{0}}{2c}L_{z}+e\phi(\vec{r})+\frac{\partial
S}{\partial t}=\frac{i\hbar}{2m}\nabla^{2}S\,,\right.  \label{p13}%
\end{align}
\noindent where the Hamilton principal function reads
\begin{align}
S(x,y,z,t)  &  =\frac{1}{2}\left[  \alpha_{1}(t)x^{2}+\alpha_{2}%
(t)y^{2}+\alpha_{2}(t)z^{2}\right]  +\xi_{1}(t)x+\xi_{2}(t)y+\xi
_{3}(t)z\nonumber\\
&  +\zeta_{1}(t)xy+\zeta_{2}(t)xz+\zeta_{3}(t)yz+\lambda_{1}(t)+\lambda
_{2}(t)+\lambda_{3}(t). \label{p14}%
\end{align}

It is worth to realize that in the limit when $\hbar$ goes to zero we obtain
the respective classical Hamilton-Jacobi equation and solution. In the
particular case where $B_{1}=0,\phi(\vec{r})=0$ and $\alpha_{3}(t)=\zeta
_{1}(t)=\zeta_{2}(t)=\zeta_{3}(t)=0,$ we shall find the Eq.(\ref{p14}) with
the Hamilton-Jacobi equation. Thus, the substitution of (\ref{p14}) into
equation (\ref{p13}) generates a polynomial equation leading to a set of
first-order ordinary differential equations.

Therefore, after resolving the corresponding set of non-linear differentials
equations, the quantum Hamilton principal function reads
\begin{align}
&  \left.  S(x,y,z,t)=-\frac{m\omega\tan(\omega t+c_{1})}{2}(x^{2}%
+y^{2})+\left(  \frac{\sigma}{m}-c_{2}\tan(\omega t+c_{1})\right)  x\right.
\nonumber\\
&  \left.  +\left(  \frac{\sigma}{m}\tan(\omega t+c_{1})+c_{2}\right)
y+i\hbar\ln[\cos(\omega t+c_{1})]-\frac{1}{2m}\tan(\omega t+c_{1})\left[
\frac{c_{2}^{2}}{\omega}+\frac{1}{\omega}\left(  \frac{\sigma}{\omega}\right)
^{2}\right]  \right. \nonumber\\
&  \left.  -\frac{c_{3}^{2}t}{2m}+c_{3}z+c_{4}+c_{5}+c_{6}.\right.
\end{align}

The solution consists in obtaining the quantum propagator if we impose the
following boundary condition \cite{b7}:
\begin{equation}
S(x,y,z,0)=\hbar k_{x}x+\hbar k_{y}y+\hbar k_{z}z. \label{p31}%
\end{equation}
Hence we obtain
\begin{equation}
c_{1}=0,\quad c_{2}=\hbar k_{y},\quad c_{3}=\hbar k_{z},\quad\sigma
=\omega\hbar k_{x},\quad c_{4}+c_{5}+c_{6}=0. \label{p32}%
\end{equation}
Now, using
\begin{align}
K(x,y,z,t;\tilde{x},\tilde{y},\tilde{z},0)  &  =(2\pi)^{-3}\int d^{3}%
k\exp\left\{  \frac{i}{\hbar}\left[  S_{k}(x,y,z,t)-S(\tilde{x},\tilde
{y},\tilde{z},0)\right]  \right\}  , \label{p37}%
\end{align}
substituting the solution which is in accordance with the initial conditions
imposed to the expression of the propagator and integrating in $k$, we arrive
to
\begin{align}
K(x,y,z,t;\tilde{x},\tilde{y},\tilde{z},0)  &  =\left(  \frac{m\omega}{2\pi
i\hbar\sin(\omega t)}\right)  \left(  \frac{m}{2\pi i\hbar t}\right)  ^{1/2}
\exp\left\{  \frac{im}{2\hbar}\left[  \omega\cot(\omega t)\left(  \left(
x-\tilde{x}\right)  ^{2}+(y-\tilde{y})^{2}\right)  \right.  \right.
\nonumber\\
&  \left.  \left.  +2\omega\left(  x\tilde{y}-\tilde{x}y\right)
+\frac{\left(  z-\tilde{z}\right)  }{t}^{2}\right]  \right\}  . \label{p38}%
\end{align}
It leads to a two-dimensional oscillator in the plane $xy$ and a free particle
in the direction $0z$.

On the other hand, the problem of an electric charge in an oscillating pulsed
magnetic field can be approached through SF. In fact, the Schr\"{o}dinger
equation reads
\begin{equation}
i\hbar\frac{\partial\psi}{\partial t}=-\vec{\mu}\cdot\vec{B}(t)\psi,
\end{equation}
\noindent where $\mu$ is a magnetic moment and is represented, according to
the reference \cite{B13}, by $\vec{\mu}=\gamma\vec{L},$ where $\vec{L}$
represents the angular momentum. Now, we perform a rotation in the reference
system where the $z$ axis is stationary, namely
\begin{align}
x  &  =\bar{x}\cos(\delta t)-\bar{y}\sin(\delta t)\,,\nonumber\\
y  &  =\bar{x}\sin(\delta t)+\bar{y}\cos(\delta t)\,,\nonumber\\
z  &  =\bar{z}\,.
\end{align}
Hence the Schr\"{o}dinger equation reads
\begin{equation}
i\hbar\frac{\partial\psi}{\partial\tau}=-\gamma\left[  \left(  B_{0}%
+\frac{\omega}{\gamma}\right)  L_{\bar{z}}+B_{1}L_{\bar{x}}\right]  \psi.
\label{p44}%
\end{equation}
For an effective static field,
\begin{equation}
B_{ef}=\left(  B_{0}+\frac{\omega}{\gamma}\right)  \hat{k}+B_{1}\hat{\imath}.
\label{p45}%
\end{equation}
Therefore, the possibility suggested by the authors of the reference
\cite{B14} is not valid for the studied system, although it is correct for a
differential equation of first-order.

Rewriting the expression of the magnetic field (\ref{p1}) only with the part
oscillating in the $x$ direction%
\begin{equation}
B(t)=B_{1}\cos(\omega t)\hat{\imath}, \label{f1}%
\end{equation}
\noindent it implies that
\begin{equation}
\vec{A}=\frac{B_{1}\cos(\omega t)}{2}\hat{k}-\frac{B_{1}\cos(\omega t)}{2}%
\hat{\jmath}\,. \label{f2}%
\end{equation}
Now, applying this result into Eq.(\ref{p4}), the Hamiltonian reads
\begin{equation}
H=\frac{1}{2m}\left(  \vec{\nabla}p\right)  ^{2}+\frac{m\gamma^{2}}{8c^{2}%
}(y^{2}+z^{2})B_{1}^{2}\cos^{2}(\omega t)-\frac{\gamma B_{1}\cos(\omega
t)}{2c}L_{x}+e\phi. \label{f3}%
\end{equation}
Using the Schr\"{o}dinger equation
\begin{equation}
i\hbar\frac{\partial\psi}{\partial t}=H\psi, \label{f4}%
\end{equation}
\noindent and the Hamiltonian given by (\ref{f3}), yields
\begin{align}
&  \left.  -\frac{\hbar^{2}}{2m}\left(  \frac{\partial^{2}\psi}{\partial
x^{2}}+\frac{\partial^{2}\psi}{\partial y^{2}}+\frac{\partial^{2}\psi
}{\partial z^{2}}\right)  +\frac{m\gamma^{2}}{8c^{2}}(y^{2}+z^{2})B_{1}%
^{2}\cos^{2}(\omega t)\psi\right. \nonumber\\
&  \left.  -\frac{\gamma B_{1}\cos(\omega t)}{2c}L_{x}\psi+e\phi\psi
=i\hbar\frac{\partial\psi}{\partial t}.\right.  \label{f5}%
\end{align}

Now we make a rotation in the coordinate system around the $x$-axis
($x=\bar{x}$) to cancel the angular momentum operator $L_{x}$. Eq.(\ref{f5})
then reads
\begin{align}
&  \left.  -\frac{\hbar^{2}}{2m}\left(  \frac{\partial^{2}\psi}{\partial
\bar{x}^{2}}+\frac{\partial^{2}\psi}{\partial\bar{y}^{2}}+\frac{\partial
^{2}\psi}{\partial\bar{z}^{2}}\right)  +\frac{m\gamma^{2}}{8c^{2}}(\bar{y}%
^{2}+\bar{z}^{2})B_{1}^{2}\cos^{2}(\omega t)\psi\right. \nonumber\\
&  \left.  -\left[  \frac{\gamma B_{1}\cos(\omega t)}{2c}+\dot{\alpha}\right]
L_{\bar{x}}\psi+e\phi\psi=i\hbar\frac{\partial\psi}{\partial\tau}.\right.
\label{f11}%
\end{align}
We choose the arbitrary angle $\alpha$ conveniently to guarantee that the
coefficient of the term $L_{\bar{x}}$ vanishes identically, implying that
\begin{equation}
\dot{\alpha}=-\frac{\gamma B_{1}\cos(\omega t)}{2c}.
\end{equation}
Substituting this value into Eq.(\ref{f11}) we can rewrite it as
\begin{equation}
-\frac{\hbar^{2}}{2m}\left(  \frac{\partial^{2}\psi}{\partial\bar{x}^{2}%
}+\frac{\partial^{2}\psi}{\partial\bar{y}^{2}}+\frac{\partial^{2}\psi
}{\partial\bar{z}^{2}}\right)  +\frac{\dot{\alpha}m}{2}(\bar{y}^{2}+\bar
{z}^{2})\psi+e\phi\psi=i\hbar\frac{\partial\psi}{\partial\tau}. \label{f14}%
\end{equation}
Now, taking the separation of variables
\begin{equation}
\psi(\bar{x},\bar{y},\bar{z},\tau)=\varphi_{1}(\bar{x},\tau)\varphi_{2}%
(\bar{y},\tau)\varphi_{3}(\bar{z},\tau) \label{f15}%
\end{equation}
yields
\begin{equation}
i\hbar\left(  \frac{\dot{\varphi}_{1}}{\varphi_{1}}+\frac{\dot{\varphi}_{2}%
}{\varphi_{2}}+\frac{\dot{\varphi}_{3}}{\varphi_{3}}\right)  =-\frac{\hbar
^{2}}{2m}\left(  \frac{1}{\varphi_{1}}\frac{\partial^{2}\varphi_{1}}%
{\partial\bar{x}^{2}}+\frac{1}{\varphi_{2}}\frac{\partial^{2}\varphi_{2}%
}{\partial\bar{y}^{2}}+\frac{1}{\varphi_{3}}\frac{\partial^{2}\varphi_{3}%
}{\partial\bar{z}^{2}}\right)  +\frac{\dot{\alpha}^{2}m}{2}(\bar{y}^{2}%
+\bar{z}^{2})+e\phi. \label{f17}%
\end{equation}

Making $\phi(\bar{x},\bar{y},\bar{z})=\phi_{1}(\bar{x})+\phi(\bar{y}%
)+\phi(\bar{z})=0$, and organizing the terms, it follows that
\begin{align}
\left[  \frac{\hbar^{2}}{2m}\left(  \frac{\partial^{2}\varphi_{1}}%
{\partial\bar{x}^{2}}\right)  +i\hbar\dot{\varphi}_{1}\right]  \frac
{1}{\varphi_{1}}  &  =Q_{1},\label{f19}\\
\left[  \frac{\hbar^{2}}{2m}\left(  \frac{\partial^{2}\varphi_{2}}%
{\partial\bar{y}^{2}}\right)  +i\hbar\dot{\varphi}_{2}\right]  \frac
{1}{\varphi_{2}}-\frac{\dot{\alpha}^{2}m}{2}\bar{y}^{2}  &  =Q_{2}%
,\label{f20}\\
\left[  \frac{\hbar^{2}}{2m}\left(  \frac{\partial^{2}\varphi_{3}}%
{\partial\bar{z}^{2}}\right)  +i\hbar\dot{\varphi}_{3}\right]  \frac
{1}{\varphi_{3}}-\frac{\dot{\alpha}^{2}m}{2}\bar{z}^{2}  &  =Q_{3}.
\label{f21}%
\end{align}
By writing
\begin{equation}
\varphi_{3}(\bar{z},\tau)=\chi_{3}(\bar{z},\tau)\exp\left(  -\frac{iQ_{3}\tau
}{\hbar}\right)  , \label{f22}%
\end{equation}
Eq.(\ref{f21}) reads
\begin{equation}
-\frac{\hbar^{2}}{2m}\left(  \frac{\partial^{2}\chi_{3}}{\partial\bar{z}^{2}%
}\right)  +\frac{\dot{\alpha}^{2}m\bar{z}^{2}}{2}\chi_{3}=i\hbar\frac
{\partial\chi_{3}}{\partial\tau}. \label{f23}%
\end{equation}
Moreover, by performing the following transformation
\begin{equation}
\chi_{3}(\bar{z},\tau)\mapsto\chi_{3}(\tilde{z},T),\bar{z}=s(T)\tilde{z},
\label{f24}%
\end{equation}
\noindent one obtains
\begin{equation}
-\frac{\hbar^{2}}{2m}\left(  \frac{\partial^{2}\chi_{3}}{\partial\bar{z}^{2}%
}\right)  +\frac{m\dot{\alpha}^{2}s^{2}\tilde{z}^{2}}{2}\chi_{3}=i\hbar
\mu\left(  \frac{\partial\chi_{3}}{\partial T}-\frac{\dot{s}}{s}\tilde{z}%
\frac{\partial\chi_{3}}{\partial\tilde{z}}\right)  , \label{f26}%
\end{equation}
\noindent where $\mu=\frac{dT}{d\tau}$.

We redefine
\begin{equation}
\chi_{3}(\tilde{z},T)=\sigma_{3}(\tilde{z},T)\exp[if(\tilde{z},T)] \label{f27}%
\end{equation}
\noindent that once substituted in the previous equation yields
\begin{align}
&  \left.  \left\{  i\mu\frac{\partial}{\partial T}+\frac{1}{2ms^{2}}%
\frac{\partial^{2}}{\partial\tilde{z}^{2}}-\frac{m\dot{\alpha}^{2}s^{2}%
\tilde{z}^{2}}{2}+\frac{1}{2ms^{2}}\left[  i\frac{\partial^{2}f}%
{\partial\tilde{z}^{2}}-\left(  \frac{\partial f}{\partial\tilde{z}}\right)
^{2}\right]  \right.  \right. \nonumber\\
&  \left.  \left.  +\mu\hbar\frac{\dot{s}}{s}\tilde{z}\frac{\partial
f}{\partial\tilde{z}}-\mu\hbar\dot{f}\right\}  \sigma+\left\{  \frac{\hbar
}{2ms^{2}}\left(  2i\frac{\partial f}{\partial\tilde{z}}\right)  -i\mu
\frac{\dot{s}}{s}\tilde{z}\right\}  \frac{\partial\sigma}{\partial\tilde{z}%
}=0.\right.  \label{f28}%
\end{align}

In addition, making $\frac{\partial f}{\partial\tilde{z}}=m\mu s\dot{s}%
\tilde{z},$ it implies that
\begin{equation}
f(\tilde{z},T)=\frac{m\mu s\dot{s}\tilde{z}^{2}}{2}+f_{T}(T), \label{f30}%
\end{equation}
\noindent where $f_{T}(T)$ is an arbitrary function of the rescaled time $T$.
Substituting this function into the equation (\ref{f28}), one can rewrite it
as
\begin{equation}
\left\{  i\mu\frac{\partial}{\partial T}+\frac{1}{2ms^{2}}\frac{\partial^{2}%
}{\partial\tilde{z}^{2}}-\frac{\mu}{2}\left[  \frac{m\dot{\alpha}^{2}s^{2}%
}{\mu}-\mu m\dot{s}^{2}+\frac{d}{dT}\left(  m\mu s\dot{s}\right)  \right]
\tilde{z}^{2}+\frac{i}{2}\mu\frac{\dot{s}}{s}-\mu\dot{f}_{T}\right\}
\sigma=0. \label{f31}%
\end{equation}

Now, we choose the arbitrary function $f_{T}(T)$ to guarantee that the two
above last terms on the left hand side are eliminated, by setting
\begin{equation}
\frac{df}{dT}=\frac{i\dot{s}}{2s}\,. \label{f32}%
\end{equation}
\noindent Integrating this equation yields
\begin{equation}
f_{T}(T)=i\ln s^{1/2}. \label{f33}%
\end{equation}
On the other hand, defining
\begin{equation}
\Omega^{2}\equiv\frac{\mu}{ms^{2}}\frac{d}{dT}\left(  m\mu s\dot{s}\right)
-\mu^{2}\left(  \frac{\dot{s}}{s}\right)  ^{2}, \label{f34}%
\end{equation}
\noindent and substituting Eqs.(\ref{f33}) and (\ref{f34}) into Eq.(\ref{f31}%
), a compact form is achieved:
\begin{equation}
\mu\left\{  i\frac{\partial}{\partial T}+\frac{1}{2\mu ms^{2}}\frac
{\partial^{2}}{\partial\tilde{z}^{2}}-\frac{ms^{2}}{2\mu}\left[  \dot{\alpha
}^{2}+\Omega^{2}\right]  \tilde{z}^{2}\right\}  \sigma(\tilde{z},T)=0.
\label{f35}%
\end{equation}

Now, making the identification
\begin{equation}
m_{0}\equiv ms^{2}\mu=const,\qquad\frac{ms^{2}}{\mu}(\dot{\alpha}^{2}%
+\Omega^{2})\equiv m_{0}\omega_{0}^{2}, \label{f36}%
\end{equation}

\noindent with $m=m_{0}$, and substituting this values into (\ref{f35}), we obtain%

\begin{equation}
\mu\left(  i\frac{\partial}{\partial T}+\frac{1}{2m_{0}}\frac{\partial^{2}%
}{\partial\tilde{z}^{2}}-\frac{1}{2}m_{0}\omega_{0}^{2}\tilde{z}^{2}\right)
\sigma(\tilde{z},T)=0. \label{f38}%
\end{equation}

For this we make the transformation $s=v^{-1}$ in (\ref{f38}) so that they can
be rewritten as%

\begin{equation}
\ddot{v}+\omega_{0}^{2}v=\frac{\dot{\alpha}^{2}}{v^{3}}, \label{f39}%
\end{equation}

\noindent and consequently we get%

\begin{equation}
\ddot{v}+\xi^{2}v=0, \label{f40}%
\end{equation}

\noindent where $\xi^{2}=\left(  \omega_{0}^{2}-\frac{\dot{\alpha}^{2}}%
{\mu^{2}}\right)  $. In this form the problem was transformed into a classical
harmonic oscillator with time-dependent frequency. We can particularize this
problem by requiring that $\xi=const$, thus obtaining the solution
\begin{equation}
v=A\cos(\eta T+\delta) ,\label{f41}%
\end{equation}

\noindent so that
\begin{equation}
\mu=\dot{\alpha}(\omega_{0}^{2}-\eta^{2})^{-1/2}, \label{f42}%
\end{equation}
and%

\begin{equation}
s=A^{-1}\sec(\eta T+\delta). \label{f43}%
\end{equation}
Therefore, it is easy to check that the conditions (\ref{f36}) are true and,
therefore, the problem is reduced when a particular case is required.

\section{Conclusion}

We studied classical and quantum solutions for harmonic oscillator-like
systems, further encompassing the driven case and with resonances as well, by
using the Hamilton-Jacobi method. For the quantum case, the propagator allows
to study the time evolution of the system, if we take into account the
Hamilton's principal function with a linear term. This term is shown to be
essential to obtain the respective quantum propagators of the systems studied.
Therefore, it can be verified that the Hamilton-Jacobi quantum formalism is an
alternative version for the quantum mechanical formulation, obtaining the
classical limit when $\hbar\rightarrow0$.

After that, we computed, through this approach, the propagator for an electric
charge in a oscillating magnetic field. Since we observed that the
Schr\"{o}dinger approach to this problem in the literature presents a
technical flaw, we computed its solutions also through the SF.

\end{document}